\documentclass[aps,pra,notitlepage,twocolumn,superscriptaddress]{revtex4-1}

\usepackage{color}
\usepackage[usenames,dvipsnames]{xcolor}
\usepackage{graphicx}
\usepackage{natbib}
\usepackage{amsmath}
\usepackage{amssymb}
\usepackage[normalem]{ulem}
\usepackage{braket}
\usepackage[utf8]{inputenc}
\usepackage{url}
\usepackage{hyperref}

\setcitestyle{super}
\begin{document}

\newcommand{\Comment}[1]{\textcolor{blue}{#1}}
\newcommand{\CommentImp}[1]{\textcolor{red}{#1}}
\newcommand{\ToDo}[1]{\textcolor{NavyBlue}{(\textbf{ToDo:} #1)}} 
%avoid clash w/ latexdiff: http://en.wikibooks.org/wiki/LaTeX/Colors

\newcommand{\eref}[1]{(\ref{#1})}

\title{Quantum walk coherences on a dynamical percolation graph}
\author{Fabian Elster}
\email{fabian.elster@uni-paderborn.de}
\author{Sonja Barkhofen}
\author{Thomas Nitsche}
\affiliation{Applied Physics, University of Paderborn, Warburger Straße 100, 33098 Paderborn, Germany}
\author{Jaroslav Novotn\'y}
\affiliation{Department of Physics, Faculty of Nuclear Sciences and Physical Engineering, Czech Technical University in Prague, B{\v r}ehov\'a 7, 11519 Prague, Czech Republic}
\author{Aur\'el G\'abris}
\affiliation{Department of Physics, Faculty of Nuclear Sciences and Physical Engineering, Czech Technical University in Prague, B{\v r}ehov\'a 7, 11519 Prague, Czech Republic}
\affiliation{Department of Theoretical Physics, University of Szeged, Tisza Lajos k\"or\'ut 84, 6720 Szeged, Hungary}
\author{Igor Jex}
\affiliation{Department of Physics, Faculty of Nuclear Sciences and Physical Engineering, Czech Technical University in Prague, B{\v r}ehov\'a 7, 11519 Prague, Czech Republic}
\author{Christine Silberhorn}
\affiliation{Applied Physics, University of Paderborn, Warburger Straße 100, 33098 Paderborn, Germany}

\date{\today}

\maketitle

{ \bf
Coherent evolution governs the behaviour of all quantum systems, but in nature it is often subjected to influence of a classical environment.
For analysing quantum transport phenomena quantum walks emerge as suitable model systems.
In particular, quantum walks on percolation structures constitute an attractive platform for studying open system dynamics of random media.
Here, we present an implementation of quantum walks differing from the previous experiments by achieving dynamical control of the underlying graph structure. 
We demonstrate the evolution of an optical time-multiplexed quantum walk over six double steps, revealing the intricate interplay between the internal and external degrees of freedom. 
The observation of clear non-Markovian signatures in the coin space testifies the high coherence of the implementation and the extraordinary degree of control of all system parameters.
Our work is the proof-of-principle experiment of a quantum walk on a dynamical percolation graph, paving the way towards complex simulation of quantum transport in random media.
}

\begin{figure*}
    \begin{flushleft}
        \includegraphics[width=183mm]{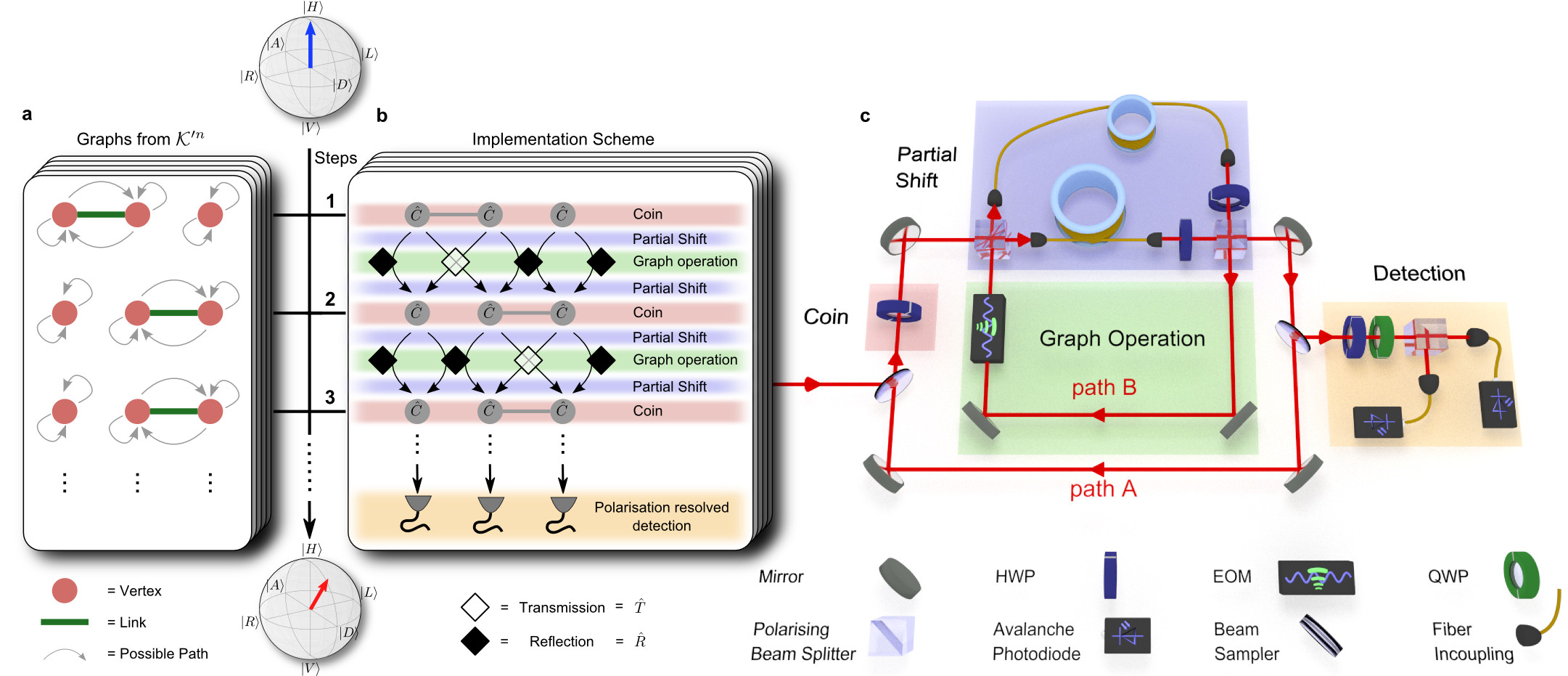}
    \end{flushleft}
    \caption{\textbf{Implementation Scheme: } \textbf{a}: One pattern of a dynamically changing graph with 3 sites used in the experiment (shown explicitly for 3 steps). The full set of patterns used in the experiment is denoted by $\mathcal{K}^{\prime n}$ for $n$ steps. The input state (blue arrow) is evolved (red arrow) and measured tomographically at every step.
\textbf{b}: Implementation scheme of the example, the $\hat{R}$ and $\hat{T}$ operators are represented by filled and hollow diamonds, respectively.
\textbf{c}: Setup scheme of the time-multiplexed PQW.
According to the implementation scheme the walker always alternates between paths $A$ and $B$. The colour coding is used to mark corresponding entities in both panels.
We average over all patterns to obtain the open system's dynamics.
			}
    \label{fig:percolation}
\end{figure*}

\begin{figure}
    \begin{flushleft}
        \includegraphics[width=89mm]{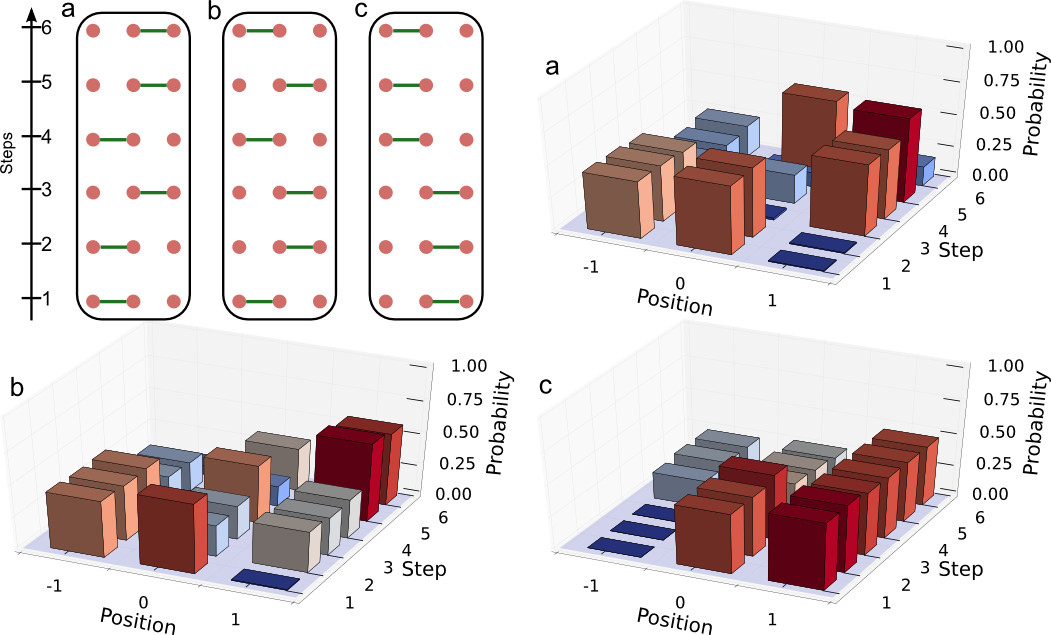}
		\end{flushleft}
     \caption{\textbf{Spatial Distributions: }Measured spatial distributions versus the step number for 3 example graph patterns (shown on the upper left panel) from $\mathcal{K}^{\prime6}$.
		Note that the height of a bar is unchanged from one step to the next if the site is disconnected. The high similarity (on average $95.6\,\%$) between the empirically observed probabilities and those from the ideal process makes a graphical comparison unnecessary. 
		}		
       \label{fig:stepdynamic}
\end{figure} 

\begin{figure*}
    \begin{center}
        \includegraphics[width=183mm]{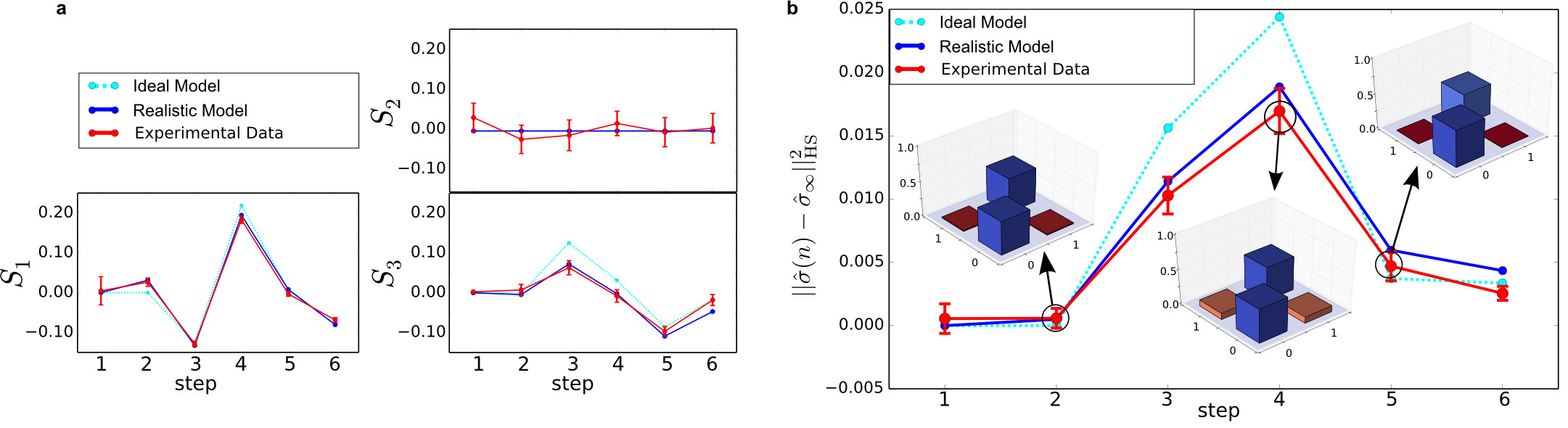}
    \end{center}
    \caption{\textbf{Coherence Measurements: }\textbf{a}: Values of the Stokes parameters of the reduced state $\hat{\sigma}(n)$ of the coin: observed (solid red lines), ideal model (dashed lines), and realistic model (blue, solid lines).
		\textbf{b}: Hilbert--Schmidt distance $||\hat{\sigma}(n)-\hat{\sigma}_{\mathrm{\infty}}||^2_{\mathrm{HS}}\equiv \mathop{\mathrm{Tr}}\left\{(\hat{\sigma}(n)-\hat{\sigma}_{\mathrm{\infty}})^2\right\}$ between $\hat{\sigma}(n)$ and $\hat{\sigma}_{\infty}$, the maximally mixed, asymptotic state: observed data (red), ideal model (turquoise), and realistic model (blue). 
		The insets show the experimental density matrix for the three chosen steps. 
		Statistical errors are smaller than the symbol size. The depicted error bars are calculated using a numerical simulation of all relevant systematic errors and are discussed in detail in the Methods section. 
		}
    \label{fig:distance}
\end{figure*} 

The development of experimentally feasible quantum simulators that are capable of supporting a wide range of phenomena is presently the target of intensive research \cite{lewenstein_ultracold_2007,johnson_what_2014-1,britton_engineered_2012}.
Discrete time quantum walks (DTQWs) \cite{meyer_quantum_1996, kempe_quantum_2003, manouchehri_physical_2013} are regarded as a promising platform for building quantum simulators.
Various theoretical studies utilise this model to analyse e.g. the occurrence of localization effects \cite{inui_localization_2004}, topological phases \cite{kitagawa_exploring_2010, asboth_symmetries_2012}, the mimicking of the formation of molecule states \cite{ahlbrecht_molecular_2012}, and even energy transport in photosynthesis \cite{mohseni_environment-assisted_2008, plenio_dephasing-assisted_2008} can be linked \cite{childs_relationship_2010}.
The high level of attracted interest and the applicability of this model system is also reflected by numerous experimental realizations of quantum walks such as in nuclear magnetic resonance \cite{du_experimental_2003, ryan_experimental_2005}, trapped ions \cite{schmitz_quantum_2009,zahringer_realization_2010}, atoms \cite{karski_quantum_2009, genske_electric_2013}, photonic systems \cite{bouwmeester_optical_1999, do_experimental_2005, broome_discrete_2010, regensburger_photon_2011, cardano_quantum_2014} and waveguides \cite{perets_realization_2008, bromberg_quantum_2009, peruzzo_quantum_2010, owens_two-photon_2011, sansoni_two-particle_2012, di_giuseppe_einstein-podolsky-rosen_2013, crespi_anderson_2013, meinecke_coherent_2013, poulios_quantum_2014}.
Using a discrete time-multiplexed quantum walk setup \cite{schreiber_photons_2010}, which provides a versatile and resource efficient system, Anderson localisation \cite{anderson_absence_1958, schwartz_transport_2007} on a one dimensional graph \cite{schreiber_decoherence_2011} and two particle interaction effects based on a two dimensional square lattice \cite{schreiber_2d_2012} have been demonstrated. 
Yet, all experiments to date have been limited to realising the walk on regular, fully connected and static graphs.
Varying graph structures \cite{chalker_percolation_1988, romanelli_decoherence_2005, kollar_asymptotic_2012} lie at the heart of percolation, which is one of the simplest yet non-trivial models at the intersection of several disciplines.
The theory of percolation is concerned with the connectivity of random graphs, which in turn can be related to the dynamics of particles, excitations or fluids propagating accross a random medium. 
The model has been extensively studied in mathematics and physics, with applications ranging from phase transitions to a multitude of transport phenomena \cite{erdos_evolution_1960, grimmett_percolation_1999, dorogovtsev_evolution_2013, steif_survey_2009}.
In the standard percolation model, links between vertices of a finite graph are present or absent with a given probability $\mathfrak{p}$.
A generalisation of the model, known as dynamical percolation \cite{steif_survey_2009}, employs a constantly changing graph, yielding a model optimized to the simulation of randomly evolving -- fluctuating medium (even space itself), network or environment.
Here, we pick up the idea of combining the percolation model with quantum mechanics, and present its experimental realization based on quantum walks on varying graph structures \cite{leung_coined_2010, kollar_asymptotic_2012}.
The resulting new device is able to simulate quantum effects in imperfect media induced by local perturbations of the graph structure.
The programmed randomness plays the role of a fluctuating external field effectively giving rise to open system dynamics exhibiting a rich diversity of subtle decoherence mechanisms.

\section{Results}
\subsection{Percolation model}
A quantum walk, defined in analogy with a random walk, is a particular quantum mechanical process on a prescribed graph, which consists of iterative applications of a unitary operator usually called a step, which factorizes as $\hat{U}=\hat{S}\hat{C}$.
The \textit{coin operator} $\hat{C}$ modifies the walker's internal coin state and is crucial for the non-trivial quantum dynamics, while the \textit{shift operator} $\hat{S}$ implements transitions across the links of the graph in dependence of the internal state.
The extension of quantum walks to dynamical percolation graph structures leads to the concept of percolation quantum walks \cite{kollar_asymptotic_2012} (PQW). 
Here, a finite set of vertices is considered, where at each step a graph with an edge configuration $\kappa$ is probabilistically chosen from all possible configurations $\mathcal{K}$. 
On the graph with configuration $\kappa$ the dynamics are defined in analogy to the DTQW. 
At the gaps, the shift operator $\hat{S}$ is modified by inserting reflection operators, leading to the shift operator \cite{kollar_asymptotic_2012} $\hat{S}_{\kappa}$. 
The probabilistic nature of the choice of the configuration $\kappa$ models an open system dynamics.
The evolution of the walker's state from $\hat{\rho}(n-1)$ to $\hat{\rho}(n)$ is described by the random unitary map (RUM)
\begin{equation}
\hat{\rho}(n) = \sum_{\kappa \in \mathcal{K}} p(\kappa,\mathfrak{p}) \left(\hat{S}_{\kappa} \hat{C}\right) \hat{\rho}(n-1) \left(\hat{S}_{\kappa} \hat{C}\right)^{\dagger},
\label{RUO_iteration}
\end{equation}
where $p(\kappa,\mathfrak{p})$ is the probability of each configuration $\kappa$.
Generally, it is assumed that open dynamics results in the gradual loss of information about the initial state, destroying all coherence. 
The system evolution under RUM contradicts this intuition and can result in a variety of non-trivial asymptotic states\cite{kollar_asymptotic_2012} attained after a dynamically rich transient regime.
The typical characteristics already manifest themselves for a graph describing a linear chain. 
For their experimental observation we needed to design an apparatus, which is able to provide the following capabilities:
first, the implementation of finite graph structures along with the dynamical creation or removal of edges between vertices;
second, the easy and quick reconfigurability of the apparatus for the collection of data over the large configuration space $\mathcal{K}$ in a short time;
third, the full access to the coin state to track coherences in the walker's state during its evolution. 

\subsection{Experimental realisation}
We based our simulator on the time-multiplexing technique \cite{schreiber_photons_2010, schreiber_2d_2012}.
Thus it inherits advantageous features such as remarkable resource efficiency, excellent access to all degrees of freedom throughout the entire time evolution, and stability sustained over many consecutive measurements providing sufficient statistical ensembles.
As before the input state is prepared by weak coherent light at the single photon level, which is appropriate for studying any single particle properties of our walk \cite{knight_quantum_2003}.
Our detection apparatus is adapted to single photon detection, which makes our interference circuit compatible for future multi-particle experiments with coincidence detection.  
The greatest challenge in this experiment has been the implementation of the dynamically changing shift operator $\hat{S}_{\kappa}$, that realises the reflecting boundary conditions as well as the dynamical creation of edges between vertices.

The implementation of the walk is based on a loop architecture where the walker is realised by an attenuated laser pulse \cite{schreiber_photons_2010, schreiber_2d_2012}. 
Its polarization, expressed in the horizontal and vertical basis states $\ket{H}$ and $\ket{V}$, is used as the internal quantum coin and manipulated by standard linear elements, performing the \textit{coin operation} $\hat{C}$.
Different fibre lengths in the loop setup introduce a well defined time delay between the polarisation components, where different position states are uniquely represented by discrete time bins (mapping the position information into the time domain). 
To attain repeated action, we have completed the apparatus with a loop geometry that consists of the two paths A and B (see Fig.~1), similarly to the 2d quantum walk \cite{schreiber_2d_2012}.
In contrast to previous experiments, here one full step of the PQW is executed by two round-trips in the loop architecture, alternating between paths A and B.
Additionally to the standard half-wave plate (HWP) in path $A$ (red area) we include a fast electro-optic modulator (EOM) in path B (green area), which now allows to actually change the underlying graph structure, and defines the additional \textit{graph operation} $\hat{G}_{\kappa}$.
It is embedded between two partial shifts $\hat{S}$ making up a full shift operation as $\hat{S}_{\kappa} = \hat{S} \hat{G}_{\kappa} \hat{S}$ thus implementing the unitary $\hat{U}_{\kappa} = \hat{S}_{\kappa} \hat{C}$.
The EOM is programmed to perform the transmission $\hat{T}$ or reflection operation $\hat{R}$ depending on whether a link is present or absent at the particular time encoded position in the configuration $\kappa$.
Thus, changing the structure or size of the graph requires only a reprogramming of the timings delivered to the EOM. 
Detection at each step by a pair of avalanche photo diodes gives us access to the time evolution in the coin as well as in the position degree of freedom.

Aiming for the reconstruction of the (reduced) density matrix $\hat{\sigma}(n)$ of the coin at the $n$th step, we perform a full tomography \cite{james_measurement_2001} of the coin state and demonstrate its evolution over six full steps.
We test our simulator by performing a PQW with a Hadamard coin on a graph consisting of three vertices and at most two links.
This choice of system size enables us to observe all relevant dynamical features within limits of the number of possible iterations due to round-trip losses.
The sample space for the complete dynamics over $n$ steps corresponds to the set $\mathcal{K}^n$ of all possible patterns of length $n$.
A restriction to the configurations $\mathcal{K}'$, obtained from $\mathcal{K}$ after removing graphs with both links present or both absent, reduces the size of the experimental sample space to $64$ for a $6$ step dynamics, while leaving the asymptotic behaviour unaffected (see supplementary material).
We realise all configuration patterns from $\mathcal{K}^{\prime 6}$ which corresponds to a link probability $\mathfrak{p}=1/2$.
The transmission $\hat{T}$ and reflection $\hat{R}$ operations realized by the EOM in the setup yield stationary asymptotic dynamics characterized by the single asymptotic state $\hat{\rho}_{\infty}$ being the identity.
The study of the distance between the completely mixed coin state $\hat{\sigma}_{\infty} = \frac{1}{2}(\ket{H}\!\bra{H} + \ket{V}\!\bra{V})$ and our measured $\hat{\sigma}(n)$ thus yields two kind of information. First, it allows us to track how far is the system from the asymptotic state, and second, any increase of the distance from the stationary state, that in our case is the completely mixed state, is a clear signature of non-Markovian evolution in the coin space.

\subsection{Finite graphs}

The individual analyses of experimental measurement results for each of the 64 patterns can be used to reveal the extent of accuracy to which the step operators $\hat{S}_{\kappa}\hat{C}$ were realized. 
Residual populations outside the positions $-1, 0$ and $1$ constituted less than $2\,\%$ on average, confirming the realization of a finite graph. %1,98%
Since an unconfined walker would have spread over a length of 12 sites over the 6 steps, the strong confinement to three sites achieved by a programmable boundary and not by a fixed one \cite{meinecke_coherent_2013} is remarkable. For horizontally polarised initial states, the experimentally obtained spatial distributions are displayed on Fig.~2 for selected configuration patterns, demonstrating the precision of the implementation of the dynamically changing graph structure.
(See the supplementary material and extended data figures for vertically polarised input.)

\subsection{Quantum percolation walk}

We implement the open system dynamics by averaging tomographic data over 64 patterns at each step $n$, corresponding to taking the average over a fluctuating external field \cite{Alicki_opensystems_2007}.
The open system dynamics is arises due to the loss of knowledge about the external field, and not due to a coupling to some external quantum heat bath.
We reconstruct the reduced coin state $\hat{\sigma}(n)$ by determining the Stokes parameters presented in Fig.~3a.
The measured parameters (red lines) are compared both to the ideal model (dotted lines) and to a realistic model incorporating the systematic errors present in the experiment (blue lines). 
All Stokes parameters are in very good agreement with the theoretical models, $S_1(n)$ and $S_3(n)$ show the oscillatory behaviour, and $S_2(n)$ is zero within the error bars.
The systematic errors lead to small deviations only in the amplitude but not in the qualitative form and oscillation periods compared to the ideal model theory. 
Details on the realistic model and the errorbars can be found in the Methods section.
On Fig.~3b we present the Hilbert--Schmidt distance \cite{buzek_quantum_1996} of our measured density matrix $\hat{\sigma}(n)$ from the completely mixed state $\hat{\sigma}_{\infty}$.

\section{Discussion}
The initially pure reduced state $\hat{\sigma}(0)=\ket{H}\!\bra{H}$ (at distance $0.5$, not shown on the plot) becomes completely mixed in a single step, however soon the distance increases.
The observed curve is part of an oscillatory evolution \cite{kollar_asymptotic_2012} that eventually decays to the maximally mixed state for the set of coin operators used in the experiment.
The non-Markovian behaviour reflected in the revival from the completely mixed state is the witness that between the position and the coin degree of freedoms sufficient coherence survives the averaging over 64 patterns.
The excellent agreement with the realistic model proves that the evolution is dominantly dictated by the controlled random unitary evolution, and other sources of decoherence, such as dephasing, are negligible.

In summary, we have demonstrated the percolation quantum walk over 6 steps using a quantum simulator exploiting enhanced time-multiplexing techniques.
As a highlight, our system is capable of realizing the walk on arbitrary, dynamically changing linear graph structures in a programmable way.
By its design the device allows access to internal and external degrees of freedom, facilitating the study of their intricate interplay, in particular revealing the exchange of coherences.
The clear revival of coherences in the coin state obtained by tomographic measurements confirms the precise control of the open system dynamics, and prove the sustained high stability of the system.

Our work paves the road to study coherence properties of systems with changing connectivity for materials resembling in structure and function grainy or porous substances.
While losses restrict our proof-of-principle experiment there is no geometric limitation on the size of the graph.
Classical light sources and amplification can be used for studying coherence properties over 300 steps \cite{wimmer_optical_2013}.
Prospective phenomena to investigate experimentally include boundary induced effects such as edge states \cite{kollar_discrete_2014} and non-trivial asymptotic behaviour, transport on percolation structures, and critical phenomena in higher dimensions.
The introduction of multiple single photon states in a system without amplifiers, will open the route for the full experimental exploration of quantum interference effects in percolated media. 

\section{Methods}

\subsection{Experimental setup.}

The laser used in the experiment is a diode laser with a central wavelength of 805\,nm. 
It produces pulses of approximately 88\,ps FWHM duration which are attenuated by several neutral density filters to a level of about 135 photons per pulse after the incoupling mirror of the experiment.
This leads to only 1.2 photons arriving at the detectors in the first step relevant to our measurements, whereas the overall round trip losses sum up to 50\,\%.
The main contributions are the coupling losses at the fibres and the losses at the incoupling and outcoupling mirrors, where we probabilistically couple 0.2\,\% into the setup and 7\,\% out at each round trip. 
The repetition rate is variable and chosen with respect to the duration of a full quantum walk. 
To realise the partial shift two single mode fibres of 135\,m and 145\,m length have been used leading to a position separation of 46\,ns.
This allows for 13 occupied positions without any overlap and signifies the maximum possible system size with this specific set of fibres.  
The EOM and its characteristics are discussed in the next section. %Section \ref{EOM}.
The detectors used are silicon-based avalanche photo diodes operating in Geiger mode with a dead time of about 50 ns and detection efficiencies around 65\,\%.
The single photon detectors were chosen since their dynamic range is more accessbile in comparison to regular photo diodes, and also as a preparation for future genuine single photon input.

\subsection{Characteristics and description of the EOM. \label{EOM}}
The operation of the electro-optic modulator (EOM) in our experiment is based on the Pockels effect. 
It has a rise time of below 5\,ns and can switch faster than 50\,ns between consecutive switchings, which is comparable to the distance between neighbouring positions in the quantum walk. 
The switched pattern can be an arbitrary non-periodic signal, however some technical restrictions apply. 
It consists of two identical birefringent crystals with their optical axes rotated relative to each other by $90^\circ$ to compensate for their natural birefringence inducing a phase $\varphi$. 
By applying a voltage $U$ an additional phase retardation $\phi_U$ can be achieved. 
The pair of crystals are rotated by $45^\circ$ with respect to the horizontal and vertical polarisation axes defined by the polarising beam splitters in our setup. The action of the EOM on arbitrary polarization states is given in the $\{ \ket{H}, \ket{V} \}$ basis by the matrix 
\begin{eqnarray}\nonumber
&\hat{G}&_\mathrm{EOM}(U) = 
R(45 ^\circ) \cdot \hat{G}_\mathrm{crystal~1} \cdot \hat{G}_\mathrm{crystal~2} \cdot R(-45 ^\circ)
\\
&=& \frac{1}{2}
\begin{pmatrix}
1 & -1 \\
1 & 1 \\
\end{pmatrix}
\begin{pmatrix}
e^{i \phi_U} & 0 \\
0 & e^{i \varphi} \\
\end{pmatrix} 
\begin{pmatrix}
e^{i \varphi} & 0 \\
0 & e^{-i \phi_U} \\
\end{pmatrix}
\begin{pmatrix}
1 & 1 \\
-1 & 1 \\
\end{pmatrix}
\\
\nonumber
&=& e^{i \varphi} \cdot
\begin{pmatrix}
\cos(\phi_U) & i \sin(\phi_U) \\
i \sin(\phi_U) & \cos(\phi_U) \\
\end{pmatrix}.
\label{eq:C_EOM_U}
\end{eqnarray}

For $\phi_U = 0$ (at $U = 0$) the EOM realises the transmission operator $\hat{T}$, and for an appropriate choice of $U$ yielding $\phi_U = \frac{\pi}2$ the reflection operator $\hat{R}$, with
\begin{equation}
\begin{split}
\hat{T} = 
\begin{pmatrix}
1 & 0 \\
0 & 1\\
\end{pmatrix},~~~~ 
\hat{R} = 
\begin{pmatrix}
0 & i \\
i & 0 \\
\end{pmatrix}.
\end{split}
\label{eq:C_EOM_T}
\end{equation}

\subsection{The realistic model and calculation of errorbars.}

We have identified four sources of systematic errors to define a realistic model of our experiment:
first, the detector and power dependent detection efficiencies, which were determined in a separate measurement; 
second, the different losses experienced in different paths due to dissimilar coupling efficiencies and path geometries, which were similarly estimated in an independent measurement with an accuracy of $\pm 2\,\%$; 
third, the transmission through the (switched) EOM is greater than $98\,\%$, but not exactly known;
fourth, the angle of the HWP defining $\hat{C}$ can be set only with a precision of $0.2^\circ$.

The power dependence of the detector efficiencies is constant from the second step on since the power in our experiment drops exponentially from step to step.
To keep the number of parameters small, the resulting correction factor for the final steps was applied as a global correction factor yielding larger errors for the first step resulting in larger errorbars.
For the determination of the parameters of the other three errors we resorted to a numerical model.
We manually varied the parameters in the ranges suggested by the corresponding independent measurement results and device specifications. 
The parameters yielding the best fit within these ranges were chosen for the realistic model presented on the figures.
The mean deviation of the statistics produced by a Monte Carlo scan of the parameters within these ranges was used to determine the size of the errorbars.
For the first step, the errorbars produced by the Monte Carlo simulation vanish due to a symmetry, leaving the aforementioned deviation of detection efficiencies as the only source of error.\\

\bibliography{Percolation_paper}

\clearpage

\end{document}